# A New Decoding Scheme for Errorless Codes for Overloaded CDMA with Active User Detection

Ali Mousavi, Pedram Pad, Farokh Marvasti
Advanced Communications Research Institute (ACRI)
Department of Electrical Engineering, Sharif University of Technology, Tehran, Iran
Email: {ali_mousavi, pedram_pad}@ee.sharif.edu and fmarvasti@sharif.edu

***Abstract*—Recently, a new class of binary codes for overloaded CDMA systems are proposed that not only has the ability of errorless communication but also suitable for detecting active users. These codes are called COWDA [1]. In [1], a Maximum Likelihood (ML) decoder is proposed for this class of codes. Although the proposed scheme of coding/decoding show impressive performance, the decoder can be improved. In this paper by assuming more practical conditions for the traffic in the system, we suggest an algorithm that increases the performance of the decoder several orders of magnitude (the Bit-Error-Rate (BER) is divided by a factor of about $400$ in some $E_b/N_0$'s). The algorithm supposes the Poison distribution for the time of activation/deactivation of the users.**

## I. Introduction

In a binary CDMA system, each user is assigned a binary signature vector. At each symbol interval, each user multiplies its symbol (which is $+1$ or $-1$) by its signature and send it through the channel. At the receiver end, the summation of the transmitted vectors embedded in noise is received (the noise almost always is assumed to be AWGN).

In under- or fully-loaded CDMA systems (where the number of users is lower than or equal to the spreading factor), it is easy to show that the optimum signature codes are Hadamard codes which are mutually orthogonal. The optimum receiver in this case is a simple correlator.

But in bandwidth limited channels, overloaded CDMA (where the number of users exceeding the spreading factor) is required. For over-loaded CDMA such orthogonal codes do not exist and the problem of finding suitable codes and decoding schemes for such systems is a challenging problem. Psudo-Noise (PN) sequences, Multiple-Orthogonal (MO) sequences [2] and binary Welch-Bound-Equality (WBE) codes [3] are examples of signature designing. Different Multi-User-Detectors (MUDs) [4] such as Parallel, Successive and Iterative Interference Cancelation [5-7] are also proposed for decoding the signals. In [8], a class of codes is proposed that guarantee errorless communication in an ideal channel and show good performance in presence of noise. A low complexity ML decoder for a subclass of proposed codes is also designed.

An assumption which is always used at the receiver end is that the active users in the systems are known. This assumption is not always valid due to the fact that the users get dynamically active and inactive in the system. Poor estimation of the active users in the system may cause the MUDs totally collapse. For more explanation, since in MUDs the data of users are extracted jointly, if the receiver incorrectly assume that an inactive user is active, tries to corresponds a $+1$ or $-1$ to that user (which sent nothing) and it affects the estimated data for the other users. Nevertheless, knowing the active users can improve the quality of the service in the system [9].

In a CDMA system, we can model the inactive users by assuming that they are multiplying zero by their signature and transmit it through the channel. Using this idea, the columns of a binary matrix that introduce a one-to-one mapping on the set of all vectors with $0$ and $\pm 1$ entries form a set of signatures that not only guarantees errorless communication in ideal channels but also has the ability of detecting active users. This set of codes was named COWDA [1]. Also, in [1] a low complexity decoder was proposed for COWDA codes and showed that it is ML in some situations. But in that paper, no assumption was presumed for the pattern of the activation/deactivation of the users in the system. In fact, the activeness of the users in the system was assumed to be a white process (there was no correlation between the activeness of any user in even consequent time samples) which is much worse than what is on hand in practical conditions. Testing under such a bad condition is a lower bound for the obtainable performance of the proposed coding/decoding scheme.

In this paper, by assuming a practical condition for the activation/deactivation pattern of the users in the system, we propose a decoding method that lowers the BER some decades. The decoder uses a usual assumption for the traffic in the system.

In the next section we explain the channel model, the used signature sequences and the traffic model of the system. In the third section we propose our new algorithm for decoding. Simulation results are presented in section IV. Section V contains a summary of the paper and some topics for future works.

## II. Channel and Traffic Model

The channel model that we use for our discussion is

$$Y = \boldsymbol{C}X + N$$

where $\boldsymbol{C}$ is the signature matrix, $X$ is the users data vector and $N$ is the Gaussian noise vector with covariance matrix $\sigma^2\boldsymbol{I}$ (the channel is assumed to be AWGN). The signature sequences that we use here are COWDA codes [1]. These codes are a subclass of COW codes [8] that provide errorless communication in ideal channels and in addition have the active user detection ability.

For the traffic model in the system we assume that the time of being active or inactive of each user is an exponential random variable with mean $\lambda$. Consequently, the times in which a user toggles its state is a Poisson process with parameter $1/\lambda$. In other words, each entry of the vector $X$ is 0 for a random interval with exponential length and takes $\pm 1$ values equiprobably for another such a random interval. In this model the ratio of the mean of the activation/deactivation time of users to the bit interval is an essential parameter. If we suppose that the bit rate of users is $R$, this parameter equals $\lambda R$. In addition, the activation/deactivation of the users and their data are independent from each other.

## III. Decoding Algorithm

In this section we first restate the proposed algorithm for decoding COWDA matrices in [1]. Then, by assuming the more practical condition that is explained in previous section, we propose a new decoding algorithm.

In order to implement ML decoder, each user should minimize $\|Y - \boldsymbol{C}\hat{X}\|_2$. Since each active user is not aware of the status of other users and only knows that its own signature contributes by a $\pm 1$ occurrence to the received vector (not 0); the user entry in $\hat{X}$ is $\pm 1$ and the rest of the entries (of $\hat{X}$) belong to the set $\{0, \pm 1\}$. Therefore the user has to choose between $2 \times 3^{n-1}$ input vectors $\hat{X}$ (with $n$ denoting the number of users). The computational complexity of this implementation of the ML decoder is tremendously high. Our decoding scheme is accomplished in three major steps. In the first step, the authors of [1] showed that the decoding problem can be reduced to a set of decoding problems with smaller code matrices. We divided the problem of decoding a CDMA system with $m = rl$ chips and $n = rk$ users to decoding $r$ CDMA systems with $l$ chips and $k$ users [1]. This results in a huge reduction of computational costs.

In the second step, the authors of [1] reduced the complexity of the smaller systems. Consider $\boldsymbol{D} = [\boldsymbol{A} \quad \boldsymbol{B}]$ where $\boldsymbol{A}$ is an $l \times l$ invertible matrix and $\boldsymbol{B}$ is an $l \times (k-l)$ matrix. The reason that $\boldsymbol{A}$ can be considered invertible is that the assumption of $\boldsymbol{C}$ being full rank is not very restricting and the columns of $\boldsymbol{C}$ can be permuted. Using this partitioning

$$Y = \boldsymbol{C}X + N = [\boldsymbol{A} \quad \boldsymbol{B}]\begin{bmatrix} X_1 \\ X_2 \end{bmatrix} + N = \boldsymbol{A}X_1 + \boldsymbol{B}X_2 + N$$

where $X_1$ and $X_2$ are $l \times 1$ and $(k-l) \times 1$ vectors, respectively. Multiplying both sides by $\boldsymbol{A}^{-1}$, we arrive at the equation:

$$\boldsymbol{A}^{-1}Y = X_1 + \boldsymbol{A}^{-1}\boldsymbol{B}X_2 + \boldsymbol{A}^{-1}N.$$

Thus the stated minimization problem becomes

$$\min_{\hat{X}_1, \hat{X}_2} \|\boldsymbol{A}^{-1}Y - (\hat{X}_1 + \boldsymbol{A}^{-1}\boldsymbol{B}\hat{X}_2)\|_2.$$

Fixing the estimate of $X_2$ (denoted by $\hat{X}_2$) the best estimate of $(X_1)_i$ which is the entry pertaining to the current user can easily seen to be

1- If $i \leq l$

$$\left(\hat{X}_1\right)_j = \begin{cases} sign\left(\left(\boldsymbol{A}^{-1}Y - \boldsymbol{A}^{-1}\boldsymbol{B}\hat{X}_2\right)_j\right) & j = i \\ softlim\left(\left(\boldsymbol{A}^{-1}Y - \boldsymbol{A}^{-1}\boldsymbol{B}\hat{X}_2\right)_j\right) & j \neq i \end{cases}$$

where $\hat{X}_2$ takes all vectors in $\{0, \pm 1\}^{k-l}$.

2- If $i > l$

$$\left(\hat{X}_1\right)_j = softlim\left(\left(\boldsymbol{A}^{-1}Y - \boldsymbol{A}^{-1}\boldsymbol{B}\hat{X}_2\right)_j\right)$$

where all entries of $\hat{X}_2$ belong to $\{0, \pm 1\}$ except for the $i^{\text{th}}$ entry of $\hat{X}$ (this corresponds to $(i-l)^{\text{th}}$ entry of $\hat{X}_2$) which only takes the values $\pm 1$.

where $softlim(x)$ acts as a soft limiter and is defined by

$$softlim(x) = \begin{cases} -1 & x < -\frac{1}{2} \\ 0 & -\frac{1}{2} \leq x \leq +\frac{1}{2} \\ +1 & +\frac{1}{2} < x \end{cases}$$

and $softlim$ acts on vectors entrywise. Thus, instead of looking between all likely estimates of $X$ (as stated earlier there are $2 \times 3^{k-1}$ such estimates) we need to only look between likely estimates of $X_2$ (there are either $2 \times 3^{k-1-l}$ or $3^{k-l}$such estimates). If $\boldsymbol{A}$ is a Hadamard matrix it can easily be shown that the above algorithm is ML. (This is because $\frac{1}{\sqrt{l}}\boldsymbol{A}$ is a unitary matrix and thus does not change the distribution of the noise).

Each user in the process of extracting its bit estimates the state of the other users that whether they are on or off. Since each active user is active for a period and each inactive user is inactive for a period, the estimate of the users' active/inactive states at the previous time sample can help us for better decoding the newly received vector. In the above decoder, for decoding any received vector, each user only uses the received vector and does not notice to the state of the users that it estimates in the previous time sample. Thus, the decoder is somehow memoryless. In the following which is the third step of decoding for decreasing the BER and is the contribution of this paper, we are going to use the estimate of a user about the activation state of the other ones for extracting its bit from newly received vector. Our decoding method has memory.

Decoding each $m$-dimensional (chip rate) received vector, we obtain an $n$-dimensional (number of users) vector with entries $0, \pm 1$. Although by using COWDA signatures the answer of the above decoding problem is unique, because of the noise in the channel the estimated vector may be not the sent one. Hence, by decoding one received vector, we may estimate the activation state of the users wrongly. For overcoming the above problem, for extracting the $t$th bit of a user, we decode from $(t-w)$th to $(t+w)$th received vectors. Using the obtained vectors, we estimate the activation state of the users at $t$th time. The decision method is that if in more than $w$ times a user is on, we suppose it is on at time $t$, and if in more than $w$ times a user is off, we suppose it is off at time $t$. If these decisions are consistent with what estimated by decoding the $t$th received vector, nothing needs to be done.

But if the estimated activation state of the $k$ users do not match with activation states that are estimated at time $t$, and they differs for more than $d$ (a constant threshold) users, then we decode the $t$th received vector again with the users' activation states that are obtained from the $2w+1$ vectors. Thus, our proposed algorithm has two main parameters, the window size $2w+1$ and the non-matched threshold $d$. It is reasonable that $w$ should be very small in comparison with $\lambda R$. Also, if we have a long block of data (longer than $2w+1$), this method can be repeated iteratively for several times or until it converges.

Notice that restricting each user to be active or inactive in decoding a received vector decrease the decoding complexity even more. Each inactive user restricts the search space for $X$ to the vectors that their corresponding entry is 0 and each active user restricts the search space for $X$ to the vectors that their corresponding entry is $+1$ or $-1$. Another worth mentioning point is that it is sufficient to replace $softlim$ by $sign$ for every user that we want to be estimated as active.

## IV. Simulation Results

To examine the behavior of the algorithm mentioned in section III we simulated a CDMA system with 64 chips and 88 users in presence of AWGN. The used signature matrix is $\boldsymbol{H}_4 \otimes \boldsymbol{C}_{16\times 22}$ where $\boldsymbol{H}_4$ is a $4\times 4$ Hadamard matrix, $\boldsymbol{C}_{16\times 22}$ is the COWDA matrix proposed in [1] and Table 1 and $\otimes$ denotes the Kronecker product. As it is explained, the decoding problem of the $64\times 88$ system is equivalent to decoding problem of four $16\times 22$ systems. Thus, we focus and discuss the decoding problem of the $16\times 22$ system.

| | | | | | | | | | | | | | | | | | | | | | |
|---|---|---|---|---|---|---|---|---|---|---|---|---|---|---|---|---|---|---|---|---|---|
| + | + | + | + | + | + | + | + | + | + | + | + | + | + | + | + | + | + | + | + | + | + |
| + | − | + | − | + | − | + | − | + | − | + | − | + | − | + | − | − | − | + | − | + | + |
| + | + | − | − | + | + | − | − | + | + | − | − | + | + | − | − | + | − | − | − | + | − |
| + | − | − | + | + | − | − | + | + | − | − | + | + | − | − | + | + | + | − | − | + | + |
| + | + | + | + | − | − | − | − | + | + | + | + | − | − | − | − | − | − | − | − | − | − |
| + | − | + | − | − | + | − | + | + | − | + | − | − | + | − | + | + | − | + | − | + | + |
| + | + | − | − | − | − | + | + | + | + | − | − | − | − | + | + | − | + | + | + | + | − |
| + | − | − | + | − | + | + | − | + | − | − | + | − | + | + | − | − | − | − | + | + | − |
| + | + | + | + | + | + | + | + | − | − | − | − | − | − | − | − | − | − | + | − | − | − |
| + | − | + | − | + | − | + | − | − | + | − | + | − | + | − | + | + | − | + | − | − | + |
| + | + | − | − | + | + | − | − | − | − | + | + | − | − | + | + | + | + | + | − | − | + |
| + | − | − | + | + | − | − | + | − | + | + | − | − | + | + | − | − | + | + | + | + | + |
| + | + | + | + | − | − | − | − | − | − | − | − | + | + | + | + | + | + | − | − | − | − |
| + | − | + | − | − | + | − | + | − | + | − | + | + | − | + | − | + | − | + | − | + | + |
| + | + | − | − | − | − | + | + | − | − | + | + | + | + | − | − | + | − | − | + | + | − |
| + | − | − | + | − | + | + | − | − | + | + | − | + | − | − | + | + | + | + | − | − | + |

Table 1. $\boldsymbol{C}_{16\times 22}$ where $+$ denotes $+1$ and $-$ denotes $-1$.

Simulations were done for $\lambda R = 1000$ and different $w$ and $d$. BER versus $E_b/N_0$ is depicted in the results of simulation. We tried to find optimum $w$ and $d$ for a given $E_b/N_0$. The final curve is compared with the previous decoder that was proposed for COWDA codes [1]. To the extent of our knowledge no coding/decoding pair exists that does not need to know the active users for proper decoding (in the overloaded case) except COWDA codes with proposed ML decoder. Hence, there are no other appropriate coding/decoding schemes for comparison. However, we have compared the COW codes with its ML decoder [8]. Obviously, this is not a fair comparison because in the COW/ML case we have considered that the receiver knows that all users are active. That is, even if a user is not active it has to send a $+1$ or $-1$ as a fill code. A COWDA/ML-MEMORYLESS and the ideal case COW/ML decoder have about $6dB$ difference in $E_b/N_0$ for same BER; but through the new decoding scheme this difference has been decreased to less than $2dB$ which is very interesting. From other point of view, in same $E_b/N_0$ the BER is decreased by several orders of magnitude.

Fig. 1 shows that longer window length results in better BER for lower $E_b/N_0$'s. It is intuitively reasonable because in lower $E_b/N_0$ situations the probability of error in decoding the received vector is higher. Thus, estimating the activation states of the users with respect to more received vectors will decrease the error probability.

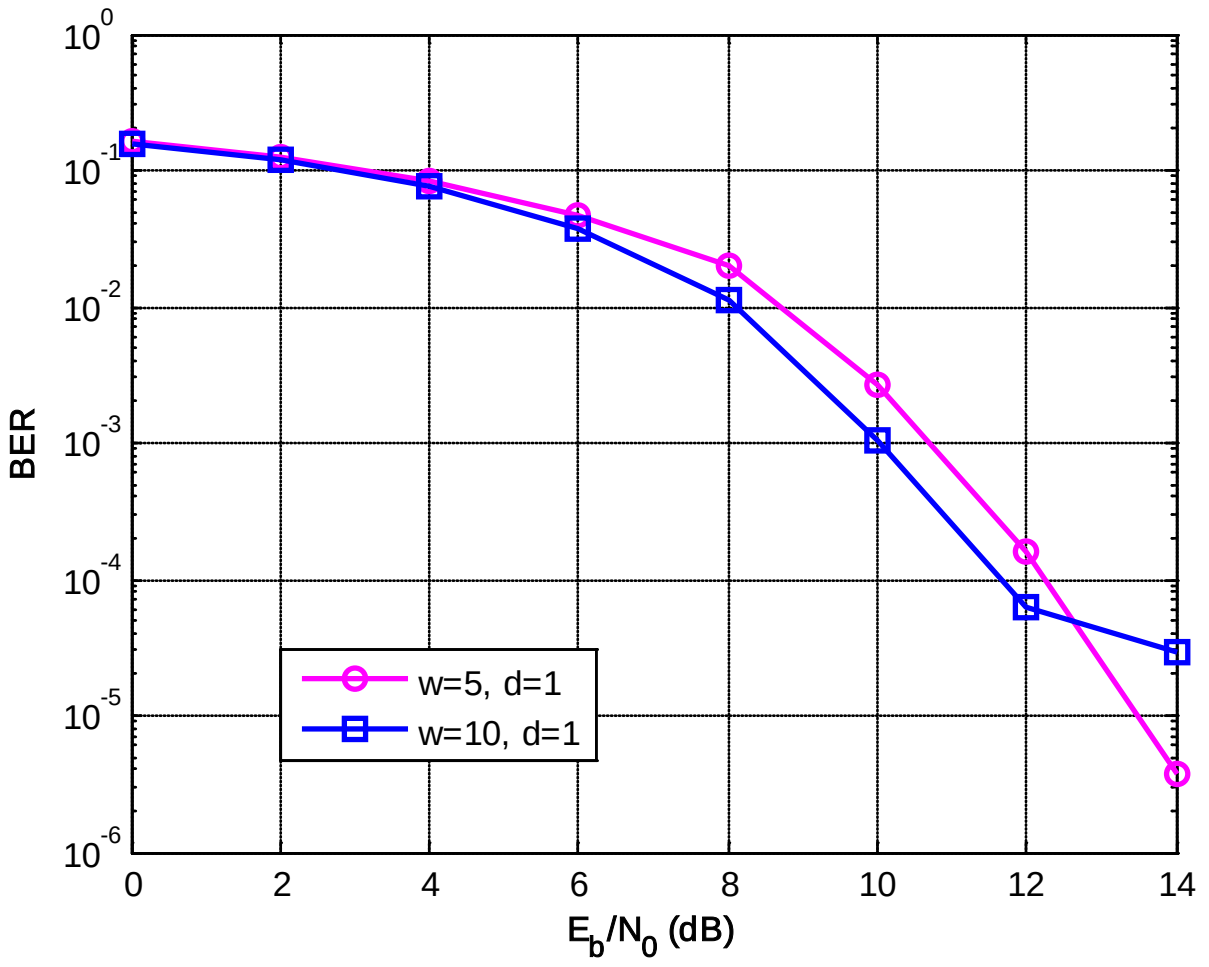

Fig. 1. BER vs. $E_b/N_0$ for different window lengths.

Fig. 2 shows the results of a simulation for finding the best error threshold ($d$). The curves show that $d=1$ is optimum for any $E_b/N_0$. This means that it is always better to find the activation states of the users by the vectors in sides of the current vector. Then using this estimation, we decode the current vector.

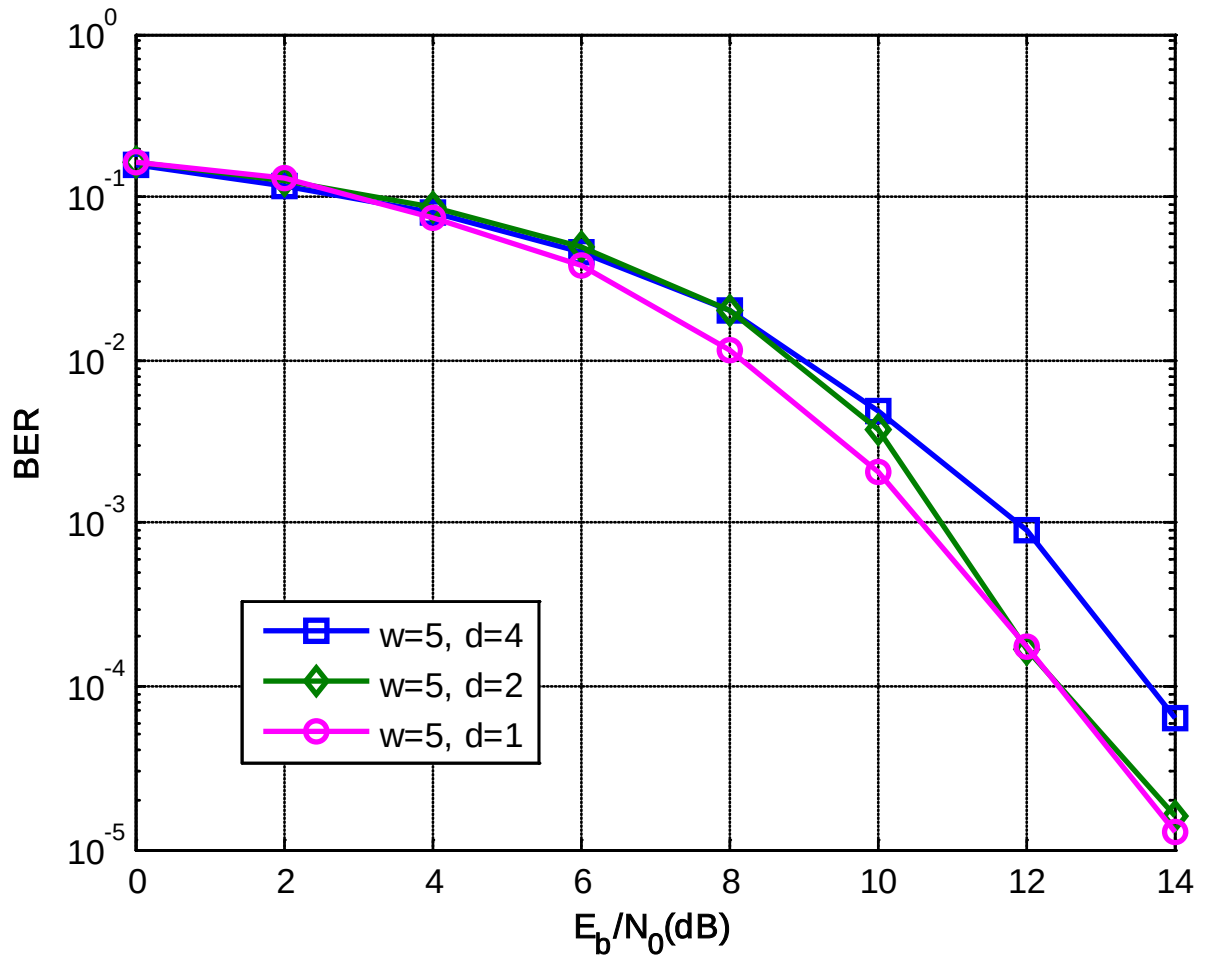

Fig. 2. BER vs. $E_b/N_0$ for different error thresholds.

Fig. 3 shows the curve of minimum BER at each $E_b/N_0$ for the simulated window lengths. This figure shows the fascinating performance of the proposed decoding method. Using this method, we can decrease the transmitted powers of the users by about $4dB$ and achieve same performance; or in other words, using same power, the BER of the system is decreased by a factor of more than $400$.

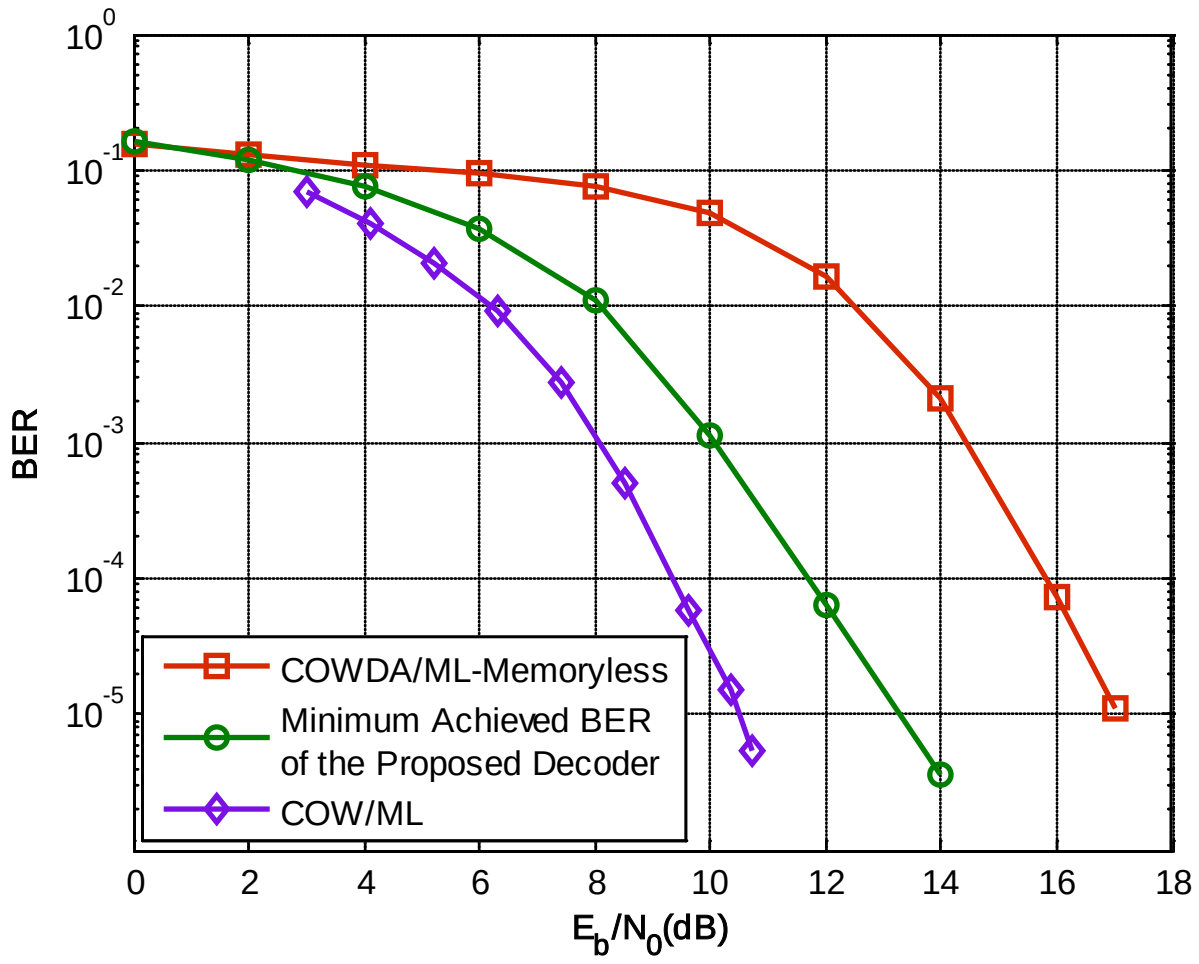

Fig. 3. BER vs. $E_b/N_0$ for different codes and decoding shemes.

## V. Conclusion and Future Works

In this paper, we have proposed a decoding scheme for a class of error-less codes that are suitable for over-loaded synchronous CDMA which are also capable of detecting active users. We mentioned that this decoding scheme is not memoryless and uses the status of all users over time to reduce the error probability. As we mentioned another important property of this algorithm is that it is ML with acceptable computational complexity. For different parameters affecting the decoding scheme, we examined different values and observed the impact of variation over curves. We concluded that for obtaining the optimized curve we should construct a combinative curve. The new decoder of COWDA is computationally feasible and simulation results indicate that this scheme is robust against additive noise. We observed that with the new scheme of decoding, for obtaining a specific BER, $E_b/N_0$ can be reduced even more than $4dB$ in comparison with the memoryless decoder. In future works, we intend to extract an analytical relationship between these parameters. Also we will work on optimizing the algorithm and finding the structure of the optimum decoder.